\documentclass[reprint, amsmath, amssymb, aps, nofootinbib,superscriptaddress]{revtex4-1}
\usepackage{graphicx}
\usepackage{dcolumn}
\usepackage{bm}
\usepackage{bbold}
\usepackage{slashed}
\usepackage{color,soul}
\usepackage{times}
\usepackage{epstopdf}
\usepackage{epsfig}
\usepackage{graphicx}
\graphicspath{ {images/} }
\usepackage{amsmath}
\usepackage[normalem]{ulem}
\usepackage[titletoc, title]{appendix}
\usepackage[colorlinks,
            linkcolor=red,
            anchorcolor=blue,
            citecolor=green
            ]{hyperref}

\begin{document}
\title{Yields of weakly-bound light nuclei as a probe of the statistical hadronization model}

\author{Yiming Cai}
\email{yiming@umd.edu}
\affiliation{Department of Physics, University of Maryland, College Park, MD 20742, USA}
\author{Thomas D. Cohen}
\email{cohen@physics.umd.edu}
\affiliation{Department of Physics, University of Maryland, College Park, MD 20742, USA}
\author{Boris A. Gelman}
\email{bgelman@citytech.cuny.edu}
\affiliation{Department of Physics, New York City College of Technology,  The City University of New York, Brooklyn, NY 11201, USA}
\author{Yukari Yamauchi}
\email{yyukari@umd.edu}
\affiliation{Department of Physics, University of Maryland, College Park, MD 20742, USA}

\date{\today}

\begin{abstract}
The statistical hadronization model successfully describes the yields of hadrons and light nuclei from central heavy ion collisions over a wide range of energies.  It is a simple and efficient phenomenological framework in which the relative yields for very high energy collisions are essentially determined by a single model parameter---the chemical freeze-out temperature.  Recent measurements  of yields of hadrons and light nuclei  covering over 9 orders of magnitudes  from the ALICE collaboration at the LHC were described by the model with remarkable accuracy with a chemical freeze-out temperature  of 156.5 $\pm$ 1.5 MeV.    A key physical question is whether (at least to a good approximation)  the freeze-out temperature can be understood, literally, as the temperature at which the various species of  an equilibrated gas of hadrons (including resonances) and nuclei chemically freeze out  as the model assumes, or whether it successfully parametrizes the yield data for a different reason.    This paper analyzes the yields of weakly-bound light nuclei---the deuteron and the hypertriton---to probe this issue.  Such nuclei are particularly sensitive to assumptions of the model because their binding energies are at a scale far below both typical hadronic scales and the freeze-out temperature.  The analysis depends only on outputs of the statistical hadronization  model, known hadronic properties and standard assumptions of kinetic theory while making no additional dynamical assumptions about the dynamics of heavy ion collisions.  The analysis indicates that a key assumption underlying the model---that hadrons (and nuclei), just prior to chemical freeze-out temperature, are in thermal equilibrium \textit{and} are sufficiently dilute as to have particle distributions accurately described statistically by a nearly ideal gas of hadrons and nuclei with masses given by their free space values---appears to be inconsistent with the chemical freeze-out temperature  output by the model, at least for these weakly-bound nuclei.   Implications of this analysis for the interpretation of parameters extracted from the model are discussed.
\end{abstract}

\maketitle

\section{Introduction}
The statistical hadronization model  (SHM) is a very simple and remarkably successful phenomenological description of the yields of stable hadrons  in central relativistic heavy ion reactions \cite{Andronic:2017pug}.    In this paper ``hadron'', when used without further explanation, connotes  light nuclei (d, ${}^3$He, the hypertriton and ${}^4$He)  as well as stable hadrons, pions, kaons, nucleons, Lambdas, etc. and unstable hadronic resonances such as $\rho$, $\omega$ and $\Delta$.  Similarly, ``stable'' indicates stability with respect to strong interactions (regardless of stability with respect to electro-weak decays). 

The model assumes that following the creation of a quark-gluon plasma, the system cools and  becomes an equilibrated hadronic gas with a volume that expands, further cooling the system. In this hadronic regime, the system is modeled as an ideal gas of various species of hadrons---both stable and unstable---following a thermal distribution at a temperature which is taken to be constant over the relevant volume.  The  masses of all hadrons are taken to be their free-space empirical values, and given the ideal gas assumption, the interactions between hadrons are assumed to be encoded solely by the existence of resonant states.  Thus, for example, pion-pion interactions are neglected except to the extent two pions can resonate into an $f_0$, a $\rho$ etc.  

The SHM assumes that as the system cools further, a chemical freeze-out temperature is reached. Beyond this point all of the stable hadrons are assumed to keep their identities, and all of the unstable hadrons are assumed to keep their identities until they decay into stable hadrons with branching ratios given by their free space values.  Thus the yields of stable hadrons (including light nuclei) are modeled as the so-called ``primordial yield'' plus the number of hadrons of the given type that come from the decay of higher mass unstable hadrons. The yields are fit by three basic parameters---the chemical freeze-out temperature ($T_{\rm cf}$), the volume of the hadronic gas at freeze-out ($V_{\rm cf}$), and the baryon chemical potential ($\mu_{\rm b}$), which accounts for differing yields of baryons and antibaryons due to the baryons in the initial state.  At high collision energies $\mu_{\rm B}$ is expected to become insignificant---and it does, empirically---and there are effectively only two free parameters.   Moreover, if one focuses on the relative yields of the different species as opposed to the absolute yields, $V_{\rm cf}$ is irrelevant; therefore, at high collision energies the relative yields effectively depend on a single parameter---the temperature $T_{\rm cf}$.

The model has proven to be remarkably successful:
\begin{itemize}
\item The yields of hadrons and nuclei are well reproduced by the SHM \cite{Abelev:2013vea,Abelev:2013xaa,ABELEV:2013zaa,Abelev:2014uua,Adam:2015yta,Adam:2015vda,Acharya:2017bso}.
\begin{itemize}
\item The yields for recent Pb-Pb collisions at $\sqrt{S_{\rm NN}} =2.76-5.02$ TeV (where the subscript NN indicates per pair of colliding nucleons) measured at the Alice detector at the LHC are well-reproduced in the model with $T_{\rm cf} =156.5 \pm 1.5$ MeV,   $\mu_{\rm b}= 0.7 \pm 3.8$ MeV and $V_{\rm cf} = 5,280 \pm 410 {\rm fm}^3 $ for $\pi^{\pm}$, $K^\pm$,$K_0^s$,$\phi$, $p$ $\overline{p}$, $\Lambda$, $\overline{\Lambda}$, $\Xi$, $\overline{\Xi}$, $\Omega^-$, ${\overline{\Omega}^+}$, $d$, $\overline{d}$, ${}^3$He, ${}^3 \overline{{\rm He}}$, ${}^3_{\Lambda}$H, ${}^3_\Lambda \overline{\rm H}$, ${}^4$He and ${}^4{\overline{\rm He}}$ \cite{Andronic:2017pug}.  (The yields of the $\Lambda$ and $\overline{\Lambda}$ include the yields of the $\Sigma^0$ and $\overline{\Sigma}^0$, which decay electromagnetically into $\Lambda$ and  $\overline{\Lambda}$ respectively and cannot be separated experimentally.)
\item These well-predicted yields cover 9 orders of magnitude.  
\item Typical predicted yields are within 20\% of the measured value for well-determined  experimental yields and within error bars of the experimental yields for less well-determined ones.  Note that  20\% errors correspond to less than 0.1 orders of magnitude, which should be compared to the 9 orders of magnitudes over which the yields range.
\item A separate fit to the light nuclei only yields $T_{\rm cf} =159 \pm 5$ MeV---which is consistent with the overall fit.
\end{itemize}
\item The model has worked well in fitting yields over a wide range of collision energies.
\begin{itemize}
\item It has successfully modeled yields from relatively low energy reactions at $\sqrt{S_{\rm NN}}=$ 2.7 -- 4.8 GeV (Brookhaven AGS) through the range $\sqrt{S_{\rm NN}}$ =6.2 -- 17.3 GeV (CERN SPS) and $\sqrt{S_{\rm NN}}=$ 70 -- 200 GeV (RHIC) to the very high energy collisions at the LHC ( $\sqrt{S_{\rm NN}} = $ 2.76 -- 5.02 TeV). 
\item The extracted chemical freeze-out temperature,   $T_{\rm cf}$, grows from $\sim 65$ MeV at the lowest energies to $\sim 155$ MeV at LHC energies. $T_{\rm cf}$ increases with increasing $\sqrt{S_{NN}}$, but saturates to $\sim 155$ MeV. $\mu_{\rm B}$ decreases with increasing  $\sqrt{S_{NN}}$ and becomes negligible by SPS energies \cite{Adams:2003xp,Andronic:2017pug,Cleymans:1992zc,BraunMunzinger:1994xr,BraunMunzinger:2003zd,BraunMunzinger:2001ip,Letessier:2005qe,Stachel:2013zma}.
\end{itemize}
\end{itemize}

 The need to understand what the model tells us about QCD has become acute with recent measurements at the LHC \cite{Abelev:2013vea,Abelev:2013xaa,ABELEV:2013zaa,Abelev:2014uua,Adam:2015yta,Adam:2015vda,Acharya:2017bso}.  The agreement of the particle yields with the model allows one to track $T_{\rm cf}$ as a function of beam energy for various experiments.   The results at the LHC imply that the model continues to work phenomenologically over a much larger range than previously seen.   The saturating behavior of $T_{\rm cf}$ can be taken as a way to decode the ``phase structure'' of QCD\footnote{Given that there is no actual phase transition at the low chemical potentials seen at the LHC, it is strictly more accurate to say that the model is being used to study the structure of the different qualitative regimes--the high T, quark-gluon plasma and the low-T hadronic gas.}. 

The key observation is that $T_{\rm cf}$ extracted by the model appears to be saturating at large beam energies, and $T_{\rm cf}$ obtained from LHC results can be regarded as the asymptotic value. It is useful to  compare the fit of the chemical freeze-out temperature,  $T_{\rm cf} =156.5 \pm 1.5$ MeV,  to estimates of the cross-over temperature---the temperature characterizing the region where a hadronic description goes over to a quark-gluon plasma description.   One way to characterize this is via the``pseudo-critical temperature'' associated with the chiral susceptibility.  Lattice studies of the pseudo-critical temperature  yield $T_c = 154 \pm 9$ \cite{Bazavov:2014pvz}  and $156 \pm 9$ MeV \cite{Borsanyi:2010bp,Borsanyi:2013bia} which, remarkably, is consistent with the extracted chemical freeze-out temperature.

This is noteworthy since it suggests that if $T_{\rm cf}$ in the model truly represents a physical chemical freeze-out,  then at sufficiently high energies the system essentially hadronizes, all hadronic species equilibrate, and then the system freezes out chemically before it cools noticeably below the cross-over temperature. Such a scenario is quite striking:  the temperature characterizing the cross-over from a quark-gluon regime to a hadronic regime  is an equilibrium thermodynamic property and is logically quite distinct from the freeze-out temperature, whose value depends on far more than equilibrium thermodynamics---it is fixed by the large-scale dynamics of the collision.

Provided that the assumptions underlying the model are valid, this remarkable scenario has strong experimental support; the SHM summarizes a significant amount of data from an extremely simple theoretical perspective.  The critical question is the extent to which this means that the simple assumptions on which the model is based are essentially correct.

The underlying basis of the model has been questioned in the past.  The principal concerns have had to do with reconciling the time scales implicit in the model with the standard understanding of the dynamics in heavy ion collisions in Ref.\cite{Heinz:2006ur,Castorina:2007eb,Schukraft:2013wba,Floris:2014pta}.  A key concern was whether there is enough time for all of the hadronic species to form and thermally equilibrate. The notion of a single chemical freeze-out temperature has also been questioned; multiple chemical freeze-out temperatures were introduced in Ref.\cite{Bellwied:2018tkc} to improve the fit. 

The approach taken in this paper is somewhat different.  We will take the model at face value and ask whether the assumptions the model makes are internally consistent in light of the experimental results, especially $T_{\rm cf}$. For the purpose of doing this we adopt a completely agnostic view on what one should expect of the dynamics of heavy ion physics.  Instead we concentrate on the properties of the equilibrated hadronic gas that is assumed to form by the model.   As will be shown, some of the model assumptions  about that gas do not appear to be internally consistent with $T_{\rm cf}$ given by the fits to experiment.   This raises critical questions about what the phenomenological successes of the model teach us about the underlying physics.  

To probe the internal consistency of the description of the hadron gas in the model we focus on the yield of weakly bound light nuclei--the deuteron (D) and hypertriton ($^{3}_{\Lambda}\rm H$). The concentration on  light nuclei is in part because much of strength of the phenomenological evidence for the success of the model rests on the yields of the light nuclei.  Of the 9 orders of magnitude in yields predicted, 5 orders of magnitude are due to the light nuclei.  
Moreover the yields of the light nuclei are, in their entirety, ``primordial'' (i.e. not from resonances) so to the extent that model is correct, they probe the equilibrium  condition prior to chemical freeze-out assumed in the model much more directly than pions, kaons, protons, Lambdas, or cascades.

The light nuclei are important for another reason: as argued in Ref.\cite{Andronic:2017pug}, the success in describing yield of light nuclei in the SHM is taken to be a signature of a statistical formation rather than due to a coalescence of baryons.  The argument is that in a coalescence picture the yield depends on the square of nuclear wave functions which vary widely between the various nuclei \cite{Csernai:1986qf,Hirenzaki:1992gx,Cho:2017dcy}.   Such a view is not universally accepted.  Variants of the coalescence model  can describe the yields of light nuclei well (for example Ref.\cite{Mrowczynski:2016xqm}) but such models require more parameters than the SHM.  This paper's focus, however, is not on the coalecense model; its sole purpose is to investigate the self-consistency of the SHM.

On the other hand, there is an {\it a priori} reason to suspect that light nuclei could be problematic in light of the model assumptions: they are all extremely weakly bound compared to the relevant scales in the dynamics associated with maintaining thermal equilibrium.  As will be seen,  these small bindings ultimately point to serious inconsistencies with key assumptions of the model.  As documentation of these issues is straightforward for the most weakly bound nuclei, this paper focuses on these.

The fact that yields of weakly-bound light nuclei are well described in the model despite violating apparently central assumptions suggests that critical assumptions could also be violated for the more plentiful hadrons despite reproducing their yields phenomenologically.  Whether this is true is critical because it goes to the heart of what one learns from this model. A key question is why the model works as well as it does phenomenologically  in spite of these inconsistencies.  The concept of partial chemical equilibrium has been studied recently in Refs.\cite{Xu:2018jff,Oliinychenko:2018ugs,Vovchenko:2019aoz} and addresses the issue.

 This paper focuses on the experiments at LHC and the analysis of them.  However the model has been applied at other $\sqrt{S_{\rm NN}}$. Weakly-bound light nuclei have been observed at lower $\sqrt{S_{\rm NN}}$, for example by the STAR collaboration \cite{Abelev:2010rv} at RHIC; this is reviewed in Ref.\cite{Chen:2018tnh}. The yields of weakly-bound light nuclei at  RHIC were studied with the SHM in Ref.\cite{Andronic:2010qu}.  In this analysis,  the model   substantially underestimates  $^3_{\Lambda}\rm{H}/^3\rm{He}$ and  $^3_{\Lambda}\rm{\bar H}/^3\rm{\bar He}$.  In this work, we will not probe the question of why the model appears to work well for the light nuclei at LHC energies but not at RHIC energies.

The next section discusses in more detail the assumptions of the model and attempts to elucidate the implication of some of these assumptions.  A particular stress will be given to the various time and distance scales that are relevant.  The following section will detail strategies for bounding the life-times of various hadrons in the supposed equilibrated gas.  Knowledge of these lifetimes is a central ingredient in testing a key assumption underlying the SHM.    Following this will be a detailed analysis of weakly bound light nuclei in a hadronic gas and a clear demonstration that they violate important assumptions underlying the SHM.  Finally there will be a discussion of the implications of these results.  

As will become clear in this analysis, the relevant binding energy need not be the total nuclear binding energy.  Rather it is the minimum energy required to separate the state into two stable constituents.  For the deuteron, this is of course the total nuclear binding of approximately 2.2 MeV.  For the hypertriton it is the separation energy into a $\Lambda$ and a deuteron. We will take this to be 0.41 MeV from the recent measurement by the STAR Collaboration \cite{Adam:2019phl}. The separation energy of hypertriton was previously taken to be $0.13$ MeV with large uncertainties. Since using the smaller value only worsens the inconsistency we will demonstrate later in the paper, we will use the recent measurement to give a conservative estimate. For simplicity of discussion, throughout this paper  we will refer to both the separation energy of the hypertriton and the nuclear binding energy of the deuteron as the ``binding energy'' and both will be denoted $B$. Natural units with $\hbar=1$,  $c=1$, $k_{\rm Boltzmann}=1$  will be used regularly in this paper.  However, following norms of the field, we will typically use MeV to denote energies or temperatures and fm to denote distances or times. For simplicity we will refer to $\Delta(1232)$ as $\Delta$ and $\Sigma(1385)$ as $\Sigma$.

\section{Assumptions}\label{assumptions}

The Statistical Hadronization Model is extremely simple.  This section discusses the assumptions that would naturally justify such a simple description of the system.   Some of these are explicit in the model.    Others are implicit but are the natural reasons why one would accept more explicit assumptions without further justification.    
The basic assumptions that would justify the model are:

\begin{enumerate}
\item The system created in relativistic heavy ion collisions achieves equilibration in a quark-gluon plasma regime.
\item The  system then expands and cools below a transition region and becomes an equilibrated hadronic gas with the bulk of the system contained in a large volume at a nearly uniform temperature \label{gas}
\begin{enumerate}
\item In this regime, the system is sufficiently dilute so that hadrons (including light nuclei)  are sufficiently well-separated  as to be discernible\label{discern}.  
\item The system is sufficiently dilute so that the relevant properties of the hadronic gas (densities of each species of hadron, and their momentum distributions as well as thermodynamic properties such as energy density and pressure) are well-approximated by a nearly ideal gas of both stable and unstable hadrons with masses given by the zero temperature value.  The interactions between hadrons are assumed to be encoded to good approximation solely by the existence of resonant states. \label{dilute}
\item The system is sufficiently dense prior to freeze-out such that interactions maintain both chemical and kinetic equilibrium for all species of hadrons.\label{interactions}
\end{enumerate}
\item As the system cools further it falls out of chemical equilibrium with the hadronic species freezing out chemically
\begin{enumerate}
\item All species of hadrons freeze out  at the same temperature to good approximation.
\item The yields seen in the detectors are given by the primordial yields given by the model for stable species at the freeze-out temperature plus yields due to the decay of unstable hadrons. Those are given by the number of those resonances, determined by the model at the freeze-out temperature, folded with branching ratios taken to be their free space values.
\item the chemical freeze-out temperature depends on the energy of the heavy ion collisions.
\end{enumerate}
\item Following chemical freeze-out, the system will remain in kinetic equilibrium with cooling temperatures until the hadronic species subsequently kinetically freeze-out and freely stream to the detector.
\begin{enumerate}
\item Unstable hadrons decay prior to reaching the detector into stable hadrons with branching ratios given by their free space values.
\end{enumerate}
\end{enumerate}

The focus of the present paper is the assumed equilibrated hadronic gas just before the putative chemical freeze-out with a principal focus on the weakly-bound light nuclei.  Before turning to the light nuclei, it is worth noting that one highly  nontrivial aspect of the SHM is Assumption \ref{dilute}.  It is by no means obvious {\it a priori} that the interactions between hadrons should be accurately encoded solely by the existence of resonant states, {\it i.e.} that a hadron resonance gas  (HRG) model should be valid.   Of course,  the HRG model will match QCD at very low temperatures, where QCD matter is a low density pion gas without substantial contributions from interactions.   Moreover, there is evidence based on lattice studies \cite{Vovchenko:2014pka,Borsanyi:2010cj,Borsanyi:2012cr,Borsanyi:2013bia,Bazavov:2014pvz} that  the HRG model does a good job in reproducing a key thermodynamic quantity ($\epsilon-3P$) of QCD at modest temperatures (up to $\sim 145$ MeV) and zero chemical potential.  But above $\sim 145$ MeV, applicability of HRG is less clear.  In the remainder of this paper we will assume that the HRG remains a viable description of the thermodynamics up to the extracted value of $T_{\rm cf}$.  It is important to recall, however, that thermodynamic quantities involve averaging and one cannot infer from a model's reasonable thermodynamic success that it has the microscopic details correct.  This is particularly true with regard to the microscopic description of rare configurations (such as nuclei) which contribute little to the thermodynamics.

This paper focuses on the description of the equilibrated matter and in particular the weakly-bound light nuclei in that matter based on Assumptions \ref{discern}, \ref{dilute} and \ref{interactions}.  It should be clear that these Assumptions, although implicit in the formulation of the model, are  essential to the description of the equilibrated matter within the statistical hadronization model. It will be shown that $T_{\rm cf}$ given by the model to describe yields implies the internal inconsistency of the model itself for the yields of light nuclei.

Note that  Assumptions \ref{discern}, \ref{dilute} and \ref{interactions} are basically the standard ones of kinetic theory \cite{Salinas:2001}.  The physical picture is quite simple: almost all of the energy in the system is contained in the mass and kinetic energy for discernible and well-localized hadrons. Thus, the energy of interaction between the hadrons is a small fraction of the total energy. Similarly, the hadrons spend almost all of their time freely propagating with their energies fixed by their masses and momenta according to the standard relativistic dispersion relation.   The hadrons  occasionally exchange energy with one another in various collisions which enables the system to establish and maintain kinetic equilibrium.   Sometimes the interactions are inelastic and change the species of one or more of the hadrons involved. This allows the system to establish and maintain chemical equilibrium.  Note that ``interactions'' in this context connotes the spontaneous decay of an unstable hadron as well as  elastic and inelastic collisions.  
\begin{center}
\begin{table}[btp]
\begin{tabular}{ |c|c|c| }
\hline
symbol&quantity \\
\hline
\hline
$n_i$ & density of hadrons of species $i$ \\
\hline
 $\epsilon_i$ & energy density of hadrons of species $i$ \\
\hline
$C_{i}$ & rate per unit volume for hadrons  \\
 & of species $i$ to be created in an interaction  \\
 \hline
 $A_i$ & rate per unit volume for hadrons  \\
 & of species $i$ to be destroyed in an interaction  \\
 \hline
 $\tau_{i}^{\rm C} \equiv \frac{n_i}{C_i}$ & characteristic time scale for the creation of \\
 &species $i$\\
 \hline
  $\tau_{i}^{\rm A} \equiv \frac{n_i}{A_i}$ &  characteristic time scale for the annihilation of \\
 &species $i$\\
 \hline
$\tau_i^{\rm int \, inel} $ & characteristic duration of an inelastic interaction that  \\ 
& creates or annihilates hadron of species $i$\\
 \hline
\end{tabular}
\caption{Some quantities characterizing interactions in a hadronic gas.}
\label{quants}
\end{table}
\end{center}

Table \ref{quants} identifies a number of quantities that characterize a putative hadron gas and the interactions that equilibrate it.  It should be clear that these cannot be defined precisely in QCD.  Consider $n_i$, the density of hadrons of species $i$. Recall that there  is no operator in QCD that measures the number of hadrons of a particular type, thus the definition of density is somewhat ambiguous from QCD perspective.   However in order for the model to make sense, the ambiguity implicit in its definition needs to be small in the sense that the scale of the ambiguities in the value of the various quantities are small compared to the value of themselves. 

It is important to clarify the meaning of $A_i$ and $C_i$, the rates per unit volume for the annihilation and creation of hadrons of species $i$. Since the SHM treats all hadrons including resonances as separate species, a hadron is considered annihilated when it undergoes an interaction changing it into another species of hadron.  Thus, for example, when a pion strikes a nucleon converting it into a $\Delta$ resonance the nucleon is considered to be annihilated.   When that $\Delta$ subsequently decays into a nucleon and a pion, a new nucleon is considered to be formed.

Of the quantities in Table \ref{quants},  the model directly  gives the densities, $n_i$ and the energy density,  $\epsilon_i$ for the various species: 
\begin{align}
n_i &=\int \frac{d^3 p}{(2 \pi)^3} \frac{g_i}{  \exp \left( \beta \left ( \sqrt{p^2 +m_i^2} - \mu_B B_i \right)  \right)\pm 1 } \label{ni} \\ \nonumber \\
\epsilon_i &= \int \frac{d^3 p}{(2 \pi)^3}   \frac{ g_i \sqrt{p^2 +m_i^2}}{   \exp \left( \beta \left ( \sqrt{p^2 +m_i^2} - \mu_B B_i \right) \right ) \pm 1 } \label{epsi}\\ \nonumber
\end{align}
where the plus sign is for fermions and the minus sign for bosons,  $g_i$ is the spin-isospin degeneracy factor of species $i$, $\beta=1/T$, $B_i$  is the baryon number of the species, 1 for baryons and -1 for antibaryons).  The (small) excluded volume factor is neglected in these expressions.

Other quantities in  Table \ref{quants} are not fixed by the SHM itself.   However, if the SHM  is correct, some of the assumptions underlying the model constrain their possible values.  

Clearly one constraint on the validity of the SHM is that  the relevant time scales characterizing the chemical equilibration---$\tau_{i}^{\rm C}$, $\tau_{i}^{\rm A}$, $\tau_i^{\rm int \, inel}$---must all be much shorter than the life-time of the fireball for all of the species of hadron, $i$ that contribute.  If this were not the case for a given species of hadrons, there would be insufficient time for that species to chemically equilibrate prior to freeze-out and thus no reason to expect $T_{\rm cf}$ to give the yield for that species. While this is worth keeping in mind when assessing the SHM, the analysis in this paper is done without any assumptions about the lifetime of the fireball or other aspects of the dynamics of the collisions and the  conclusions we reach do not depend on any knowledge of the dynamics.

In order for the model to make sense, there are constraints on these quantities relating to properties of the putative equilibrated hadronic gas just prior to chemical freeze-out:
\begin{equation}
\tau_i^{\rm A} =\tau_i^{\rm C} \equiv \tau_i  \; \; {\rm for \; all \; species \; i} \label{eqtimes}
\end{equation}

This constraint follows trivially from the assumption of equilibration.  By definition, in equilibrium, the rate at which a hadron of any species is created is identical to the rate at which it is destroyed, thus $A_i=C_i$ and Constraint (\ref{eqtimes}) follows given the definition of $\tau^{\rm A}_i$ and $\tau^{\rm C}_i$.  Intuitively $\tau_i \equiv \tau_i^{\rm A} =\tau_i^{\rm C} $ may be thought of as the characteristic lifetime of a hadron of species $i$ in the medium.  That is,  $\tau_i$  gives the typical time between when a particular hadron of type $i$ is created and when it is destroyed.

There are other important constraints that also need to be satisfied and which pose serious tests of the assumptions underlying the model:
\begin{align}
 2 \tau_i^{\rm int \, inel} &\ll {\tau_i } \; \; {\rm for \; all \; species \; i} \label{ineq1}
\end{align}

Constraint (\ref{ineq1})   encodes the need for particles to spend most of their time freely propagating with their energies fixed by their masses and momenta according to the standard relativistic dispersion relation, and with their momenta distributed (to good approximation) according to a thermal distribution for a non-interacting gas. This constraint is required for Assumption \ref{dilute} to hold.    As noted above,  there is an intrinsic ambiguity in defining the number of particles of any particular type in the gas and the model is sensible only to the extent that such an ambiguity is small.  Thus, for example, if one considers a process in which a pion plus a Lambda goes to a nucleon plus a kaon, the number of pions is one smaller after this process as compared to before,  as is the number of Lambdas.  Similarly, the number of kaons and nucleons are each one larger after the process as compared to before.  However, while the process is ongoing it is not clear how many of each of these hadrons exist--there is an ambiguity.  Constraint  (\ref{ineq1}) puts into mathematical form the statement that for the model to make sense, the ambiguity in the density of particles of a given species must be small compared to the density itself.   Note that there is no reason for the  equilibrium phase-space density to yield (nearly)  ideal gas results independent of the detailed mechanism of creation and annihilation unless this condition is satisfied.  The factor of 2 on the right-hand side of Inequality (\ref{ineq1}) encodes the fact that time is required both to create the hadron before it propagates and to destroy it subsequently, and that there is ambiguity in the number of hadrons of a given type involved in the reaction during both processes.   The time for the process to destroy the hadron should be the same as the time to create it since this is the inverse process and the factor of 2 follows.


Minimally,  the Inequality (\ref{ineq1}) 
needs to be satisfied for a species $i$ for  the model to be valid for that species. Thus, to check the consistency of the model it is important to determine the values of  $\tau_i$ and  $\tau_i^{\rm int \, inel}$---or at least constraints on their values.

As it happens,  not only is it impossible to determine $\tau_i^{\rm int \, inel}$  directly from the SHM, it also cannot be determined within the framework of kinetic theory.  Indeed, Inequality  (\ref{ineq1}) needs to be satisfied for kinetic theory itself to be applicable in the first place.  Fortunately, as will be discussed in Sect.~\ref{tests},  there is a very simple argument based on causality that sets a stringent---and very conservative---upperbound on $\tau_i^{\rm int \, inel}$ when $i$ is a light nucleus.

\section{Bounds on  $\tau_i$ for nuclei in medium \label{taui}}

The purpose of this section is to provide an upper bound on $\tau_i$  in medium for light nuclei.  Let us focus on the most weakly bound of these: the deuteron (D) and the hypertriton (${}_\Lambda^3$H). To a very good approximation,  in the regime of interest,  these can be considered as nonrelativistic  bound states containing two weakly-bound constituents: the deuteron as a proton-neutron bound state (with a binding energy of 2.22 MeV) and the hypertriton as a bound state of the deuteron and a $\Lambda$ (with a binding energy of approximately 0.41 MeV \cite{Adam:2019phl}).

The key result of this section is that for a putative equilibrated gas of hadrons and nuclei at $T=156.5$ MeV,
\begin{equation}
\begin{split}
\tau_{\rm D}&< 1.2 \, {\rm fm} \\
\tau_{{}_\Lambda^3\rm H} &< 1.0 \, {\rm fm} \; .
\end{split}
\label{taunuc}
\end{equation}

This result depends  on the assumptions underlying the SHM plus the input parameters for the model along with the assumption that interactions between pions and nucleons in weakly-bound nuclei are well-approximated by the interactions in free space.  

One does not need detailed cross-section information to obtain this result (as one would for a full kinetic theory calculation): the only information needed beyond that which enters the SHM are the well-determined  lifetimes of two resonances: $\Delta$ and $\Sigma$. The $\Delta$ and $\Sigma$ lifetimes play a key role in determining bounds on the lifetime of nucleons and $\Lambda$ baryons in medium (i.e. the characteristic time the a nucleon (Lambda) survives in medium before converting into another baryon).  A bound on the lifetime of the light nuclei turns out to depend on those lifetimes.  A method to extract the lifetime of  nucleons and Lambdas is described in the next subsection.

\subsection{Hadron lifetimes in medium}
  
 In principle, $\tau_i$ for hadrons can be obtained from a kinetic theory description  if the assumptions underlying the SHM are valid.    In particular, provided one knows all of the relevant reaction rates, kinetic theory allows one to compute  $A$ and  $C$ (which are equal in equilibrium)  from which  $\tau_i$ follows.  It should be noted that in the context of kinetic theory, the typical lifetime of a hadron of species $i$ will correspond to the relaxation time---the exponential time constant characterizing how a system with a small  excess  or shortage  of species $i$ returns to equilibrium---provided the rate of the dominant mechanism creating species $i$ is independent of the density of particle of type $i$ and the rate of the dominant mechanism annihilating species $i$ is linear in its density.  In other cases one expects that the relaxation time will be on a similar scale but differing by a factor of order unity.
 
 Unfortunately any attempt to implement  a full kinetic theory description of the system will be complicated by the fact that kinetic theory calculations require knowledge of the cross-sections of the various processes involving all of the hadrons in the system while, in the context of the SHM, such hadrons include unstable ones. This means, for example that to fully implement a kinetic theory description, one would need to know the cross-sections for reactions involving the  unstable particle (e.g. the cross-section for $\pi+ \Delta \rightarrow N + \rho$) for which there is no direct experimental data. 
 
To evade this practical problem, we adopt the following strategy:  rather than attempt to compute  $\tau_i$ itself, we will attempt to establish an upper bound on it.   Using an upper bound for $\tau_i$ still allows a meaningful test with Condition (\ref{ineq1}), albeit a less stringent one than using  $\tau_i$ itself: for the SHM to be valid, Condition (\ref{ineq1}) needs to hold using an upper bound  for $\tau_i$ in place of $\tau_i$---since if that test fails, Condition (\ref{ineq1}) will not be satisfied.

For hadrons that are unstable in free space it is trivial to bound $\tau_i$.  Their annihilation rate has two parts, one comes from spontaneous decays as in free space and the other from collisions. In a kinetic theory description these two effects are independent:
\begin{equation}
A_i =A_i^{\rm collision}  + A_i^{\rm decay} > A_i^{\rm decay} 
\end{equation}
where the inequality follows from the fact that $A_i^{\rm collision} $ is nonzero and positive. Since $\tau_i=n_i/A_i$, a knowledge of a lower bound of $A_i$ translates directly into an upper bound on $\tau_i$, and one can obtain the lower bound of $A_i$ by determining $A_i^{\rm decay}$. This does not involve a detailed kinetic theory calculation. $A_i^{\rm decay}$ can be determined from the lifetime of the resonant state in its rest frame (parameterized via its inverse, the width of the resonance $\Gamma_i$) weighted  in the distribution by  $1/\gamma = m_i/\sqrt{m_i^2 +p^2}$ where $\gamma$ is the  time dilation factor.  Thus, 
\begin{equation}
\begin{split}
A_i & > n_i \Gamma_i \langle 1/\gamma \rangle 
	> \frac{n_i\Gamma_i}{\langle \gamma \rangle} 
    = n_i \Gamma_i\frac{n_i m_i}{\epsilon_i}
\end{split} \label{Aibound}\end{equation}
where the second inequality  follows from an application of the Cauchy-Schwarz inequality. 
Thus $\frac{\epsilon_i}{\Gamma_i n_i m_i}$ is an upper bound on $\tau_i$ for a resonant state within the model.

Remarkably, if one accepts the assumptions of the SHM, one can also bound $\tau_i$ from above for hadrons that are stable under strong interactions.  The key point is that one can use either $C_i$ or $A_i$ to determine $\tau_i$ and whenever a resonance decay yields a stable hadron as a decay product, in the context of this simple model it is considered as creating that stable hadron.  Thus for a stable hadron   
\begin{equation} C_i =C_i^{\rm decay} + C_i^{\rm collision} > C_i^{\rm decay} 
\end{equation}
where the superscript ``decay'' indicates that the hadron is created via the spontaneous decay of a resonant state.  Thus,
\begin{equation}
\begin{split}
 C_i >C_i^{\rm decay} &= \sum_{j} \sum_{a}   A^{\rm decay}_j \,b_j^a  N_{i}^a >  A^{\rm decay}_k \,b_k^c  N_{i}^c\\
&> b_k^c  N_{i}^c \frac{\Gamma_k \
 n_k^2 m_k}{\epsilon_k} \end{split}
\end{equation}
where $j$ represents a possible resonance that decays into $i$ and $a$ the decay mode of resonance $j$, $b_j^a$ is the branching ratio for resonance $j$ to decay in channel $a$ and $N_{i}^a$ is the number of stable hadrons of type $i$ created in decay channel $a$; $k$ represents a particular species of resonance (typically chosen to be the one with the largest contribution) and $c$ a particular decay channel for resonance $k$.  The second inequality follows from the fact that contributions from neglected resonances are manifestly nonnegative and some are nonzero, and the third inequality again follows from the application of the Cauchy-Schwarz inequality. Since $\tau_i \equiv n_i/C_i$ it follows that 
\begin{equation}
\tau_i <\left( \frac{n_i }{n_k }\right ) \left (\frac{\epsilon_k} { n_k m_k} \right ) \left(\frac{1} {\Gamma_k  b_k^c  N_i^c} \right ) \; 
\label{taubounde}\end{equation} 
the first factor is the ratio of the densities of the stable hadron of interest to the hadronic resonance whose decay creates the particle; the second factor is the average time dilation; the third factor is the lifetime of the resonance  appropriately modified by the branching ratio and the number of stable hadrons of interest produced in the decay channel.  One should note that the branching ratio and number of hadrons of each type produced in decays are used explicitly in the SHM.  The only additional information needed is resonance width, $\Gamma_k$ which we will take from the PDG \cite{Tanabashi:2018oca}.

The power of Inequality (\ref{taubounde}) is illustrated in a bound on the lifetime of the nucleon in a medium satisfying the assumptions of the SHM with the parameters extracted from the recent LHC run  ($T_{\rm cf}=156.5$ MeV and $\mu_B \approx 0$).  We do this by taking the resonance $k$ to be the $\Delta$ since in the SHM $\Delta$ decays provide a substantial fraction of the nucleons observed at the detectors.  Since there is only one decay mode of the $\Delta$ , $\Delta \rightarrow N \pi$ and that decay yields a single nucleon, $b_\Delta^c$ is unity as is $N_{N}^c$.  Inequality (\ref{taubounde}) then implies
\begin{equation}
\tau_{N} < 2.4 \,\,\rm fm \label{tN} \; .
\end{equation}
Assuming that the hadrons get into equilibrium as assumed by the model, this implies that if a nucleon were to be dropped intact into the equilibrated medium (with a probability having momentum $p$ given by the equilibrium distribution)  within a time of 2.4 fm, one could expect a nucleon to be struck by a pion and convert into a $\Delta$ resonance thereby destroying it as a nucleon.   Of course, since the preceding calculation dealt with the creation of the nucleon by the decay of a $\Delta$, not its destruction via the creation of a $\Delta$, but by detailed balance, the rates must be the same in equilibrium.

An analogous calculation bounds the lifetime of the $\Lambda$ in medium, using spontaneous decays of the $\Sigma$ which decays into the $\Lambda$ baryon with a branching ratio of 0.87 \cite{Tanabashi:2018oca}:
\begin{equation}
\tau_{\Lambda} < 5.3 \,\,\rm fm \label{tLam} \; .
\end{equation}

\subsection{Weakly-bound light nuclei lifetimes}

The primary focus of this paper is on  weakly-bound light nuclei.  One cannot directly use Inequality (\ref{taubounde}) to get a bound on $\tau_i$ when $i$ is a nucleus, since the primary way that nuclei are created in such a medium is not through the decay of a heavy resonance with baryon number greater than one.  However, Inequality (\ref{tN}) can provide a very strong, albeit indirect, bound: when a nucleon inside a nucleus is destroyed by becoming a $\Delta$ the nucleus itself is destroyed.  

  The deuteron and hypertriton are sufficiently weakly bound that the process that annihilates  a nucleon (Lambda baryon) via pion  absorption to create a $\Delta$ (or $\Sigma$) should be well approximated by the analogous process for a nucleon (Lambda) in free space.

One expects that a  scattering process off a constituent will be  approximated well by the free-space scattering process provided: i) the binding energy is negligibly small compared to the scale of the energy transfer and other relevant scales and ii) the two components are sufficiently far apart so that the  incoming particle (in this case a pion) has sufficient momentum to clearly resolve the separate constituents. Condition i) is clearly satisfied here.   Condition ii) requires that in the rest frame of the composite system, $p_{\rm external} R \gg 1$, where $p_{\rm external}$ is the momentum of the incoming particle (a pion in this case) and $R$ is the typical spatial size of the quantum mechanical wave function for the relative separation of the constituents.  For our case the momentum of the pion necessary to create a $\Delta$ ($\Sigma$) off of a nucleon (Lambda) is $\sim 300$ MeV   ($\sim 260$ MeV).   Weak binding implies that the quantum mechanical wave functions for the relative position of the two components is dominated by regions where they are far apart. In the limit where the strength of the interaction is tuned so that the energy goes to zero while the range of the interaction is held fixed, the RMS  separation goes to $\frac{1}{\sqrt{4 B \mu}} $ where $B$ is the binding energy and $\mu$ the reduced mass.  Thus as the binding energy approaches zero, the typical separation of the constituents diverges as $B^{-1/2}$.  In fact, $\frac{1}{\sqrt{4 B \mu}} $  is a conservative estimate for the size of the state:  for ground state wave functions dominated by nodeless s-waves with  at most two points of inflection for $u =r \psi$ (as one has for typical potentials that are repulsive at short distance and attractive at longer distances), $\frac{1}{\sqrt{4 B \mu}} $  will  always underestimate the RMS size of the wavefunction.  Thus a conservative estimate for  Condition ii) is that $\frac{p_{\rm external}}{\sqrt{ 4 \mu B}} \gg 1$.  This condition is satisfied rather well for both the deuteron ($\frac{p_{\rm external}}{\sqrt{ 4 \mu B}}  \approx 5$ ) and the hypertriton ($\frac{p_{\rm external}}{\sqrt{ 4 \mu B} }\approx $ 8) so it is reasonable to approximate  scattering processes off of constituents by the analogous free-space process.

\begin{figure}[h]
    \centering
    \includegraphics[width=0.45 \textwidth]{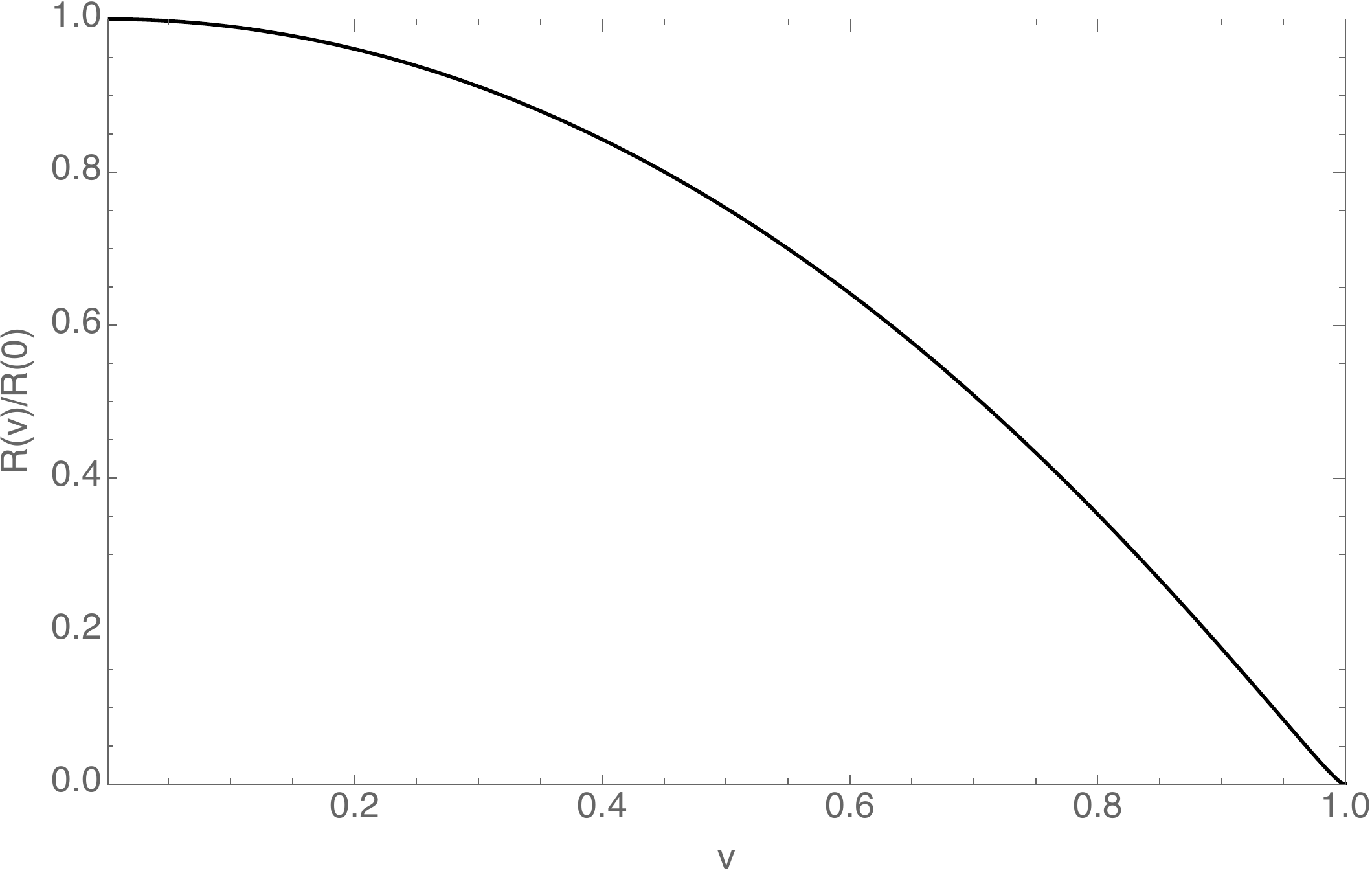}
    \caption{The ratio of the rate that nucleons with velocity $v$ absorb pions to become $\Delta$s to the rate at zero velocity for an equilibrated pion gas at 156.5 MeV as computed from Eq.~(\ref{R}). }
    \label{Rv}
\end{figure}

Given this approximation, one can determine an upper bound for the lifetime of the nuclear state.  In the context of the model, the nucleus will be considered annihilated if either of its constituents is annihilated  by being converted to another hadron, or if the constituents are knocked out in a collision. Thus the annihilation rates of constituents give a lower bound to the total annihilation rate of the nucleus. Each constituent $j$ decays with a rate of  $\frac{1}{\tau_j^{\rm bound}}$ with the superscript ``bound'' indicating that the lifetime of  the nucleon in the hadron gas may differ from the lifetime of that same species in free space for calculable kinematic reasons.  The interactions of the pions in the hadronic gas with nucleons  (Lambda baryons) depends on the velocity distribution of the nucleons (Lambdas)  and the  distribution is different for free nucleons (Lambdas) equilibrated in the gas and those bound  in deuterons (hypertritons) equilibrated in the gas.    Thus, the rate of annihilation of  the nucleus is bound by  
\begin{equation}
 A_{\rm nucleus} > \sum_{j=\rm constituents} \frac{n_j} {\tau_j^{\rm bound}}
 \end{equation}
From which it follows 
 \begin{equation}
   \tau_{\rm nucleus} = \frac{n_{\rm nucleus}}{A_{\rm nucleus}} < \frac{1}{ \sum_{j=\rm constituents} \frac{1}{\tau_j^{\rm bound}}}   
 \label{constit}\end{equation}
where $n_j$ is $n_{\rm nucleus}$ times the number of constituents $j$ per nucleus. Using the logic of kinetic theory, it is straightforward to show that the rate that nucleons of velocity $v$ are converted to $\Delta$ is given by 
 \begin{equation}
 R(v)  =C {\int_{|\phi_-(v)|}^{|\phi_+(v)|}  dp \frac{p^2}{\exp \left(\beta \sqrt{p^2 + m_\pi^2} -1 \right) \left(\gamma p v \right)}} \label{R}
\end{equation}
where $\phi_{\rm}(v)$ are the two roots of
\begin{eqnarray}
 &\gamma\sqrt{\phi_\pm(v)^2+m_{\pi}^2}+ \gamma v \phi_\pm(v) - \sqrt{p_0^2+m_{\pi}^2} =0\nonumber\\
 &\left(\sqrt{p_0^2+m_{\pi}^2}+m_N \right)^2 = m_{\Delta}^2+p_0^2\nonumber
\end{eqnarray}
where $C$ is a constant containing information about the cross section as well as constant factors and $\gamma=1/\sqrt{1-v^2}$. There is an analogous expression for $\Lambda$s converting to $\Sigma$ .  The key point for the parameters relevant for our problem is seen  in Fig.~\ref{Rv}: $R(v)$ drops monotonically as $v$ increases from zero towards unity.

This monotonic decrease of the decay rate with $v$ means that if the bound nucleons (Lambdas) in the gas have a probability distribution of speeds that is strictly slower than the velocity distribution of equilibrated free  nucleons  (Lambdas) then the average annihilation rate into $\Delta$ ($\Sigma$) per particle will be larger than for free space provided that the assumption that the scattering process off weakly bound composites is essentially the same as in free space.   This in turn means that the $\tau_i^{\rm bound}$ is smaller than the limit placed on $\tau_i$ due to pion absorption where $i$ is a nucleon or $\Lambda$.  (At a logical level, this need not imply that $\tau_i^{\rm bound}<\tau_i$ but this is unnecessary for our purposes).   In this context a strictly slower velocity distribution is one that satisfies the condition that
\begin{equation}
\begin{split}
&{\rm whenever}  \; \; \int_0^{v^{\rm bound}} dx P^{\rm bound}(x) = \int_0^{v^{\rm free}} dy P^{\rm free}(y) \\
& v^{\rm bound}<v^{\rm free}
\end{split}
\end{equation}
where $P^{\rm bound}(v^{\rm bound})$ is the probability distribution for speeds of the nucleons bound in the nucleus and $P^{\rm free}(v^{\rm free})$ is that in free space.  Since the internal motion of the constituents in these weakly bound states are very nonrelativistic, the velocity distributions of the constituents are essentially the velocity distributions of the nuclei that contain them. If, as the model assumes, the velocity distributions  for the nuclei and the nucleons are the standard ideal gas expressions, the velocity distributions for the nuclei are strictly slower than for the nucleons which implies the distribution for bound nucleons (Lambdas) is strictly slower than for free nucleons (Lambdas).   Given Eqs.~(\ref{tN}) and (\ref{tLam}), this implies that 
\begin{equation}
\begin{split}
\tau_N^{\rm bound} & <  2.4 \, {\rm fm} \\
\tau_{\Lambda}^{\rm bound} & <  5.3 \, {\rm fm} \\
\end{split}
\label{taubound}
\end{equation}

The principal result of this section, Inequality (\ref{taunuc}), follows immediately if one inserts Inequality (\ref{taubound}) into Inequality (\ref{constit}).

 The most significant point about the bound obtained in Inequality (\ref{taunuc}) is the small size of the lifetime: for T=156.5 MeV it is  of order one fermi or less.  This bound derived from basic assumptions of the SHM will turn out to be small enough that it will prove inconsistency with basic assumptions of the SHM.  

One can independently verify that the very small scale obtained as an upper bound for the lifetime of the deuteron equilibrated at T=156.5 MeV is reasonable by doing a limited kinetic theory calculation including only scattering processes in which pions dissociate deuterons (which includes processes where $\Delta$s are created and subsequently decay).  Such a calculation will yield an upper bound for $\tau_{\rm D}$ since it only includes a subset of process that can destroy a deuteron.  A calculation using empirical  inelastic scattering data yields an estimated upper bound for $\tau_{\rm D}$ quite similar to the bound in Eq.~(\ref{taunuc}).  We prefer to use the upper bound in Eq.~(\ref{taunuc}) as it does not require additional empirical input from scattering data.  However, a kinetic theory estimate does indicate that the bound obtained indirectly is of the scale that one should expect.

\section{Yields of weakly-bound  light nuclei and the self-consistency of the statistical hadronization model \label{tests}}

This section discusses whether the chemical equilibration of deuterons and  hypertritons in the putative hadronic gas at 156.5 MeV is consistent with the assumptions underlying the SHM outlined in Sec.\ref{assumptions}.  We find large violations of these assumptions. 
In particular we find compelling evidence that   Inequality (\ref{ineq1}) is  badly violated for both $\rm D$ and $^{3}_{\Lambda}\rm H$. This rules out the possibility that the yields of these nuclei are due to a chemical freeze-out of a nearly ideal gas of these particles at $T=156.5$ MeV.

Tests of Inequality (\ref{ineq1}) 
 depend on a knowledge of $ \tau_{\rm D}^{\rm int \, inel}$ and 
$\tau_{{}_\Lambda^3 \rm H}^{\rm int \, inel}$, the duration of the processes that produce or destroy a deuteron and hypertriton, respectively.  We have a reliable and extremely conservative method based on causality to obtain upper bounds on  $ \tau_{\rm D}^{\rm int \, inel}$ and 
$\tau_{{}_\Lambda^3 \rm H}^{\rm int \, inel}$.  
The basic idea is that whatever process is reasonable for creating a self-bound object that is static in its own rest frame must be long enough for the various parts of the object to be causally connected with each other given the fact that no information travels faster than the speed of light.  For  nucleus $i$, this sets  $ \frac{R_i}{\gamma_i}$  as a lower bound for $ \tau_i^{\rm int \, inel}$  where  $R_i$ is the characteristic size of the nucleus in its rest frame and $\gamma_i =\frac{ \epsilon_i}{m_i n_i}$ is a time-dilation factor accounting for the fact that the nucleus is moving.  (These $\gamma$ factors are close to unity since the mass of the nuclei is much larger than the temperature.) We can reasonably take $R_i$ to be the rms separation of the constituents making up the state as a conservative estimate for the size---it is conservative as it does not include the size of the constituents themselves. Moreover in the previous section we noted that for the  weakly bound systems considered here, $\frac{1}{\sqrt{4 B_i \mu_i}} $ (where $B_i$ is the binding energy and $\mu_i$ is the reduced mass of the two constituents in the { nonrelativistic} system) is a conservative  estimate for $R_i$.  

Thus it follows that for weakly bound nuclei describable as two-component bound states
\begin{equation}
\frac{m_i n_i}{\epsilon_i\sqrt{4 B_i \mu_i}}  \ll \tau_i^{\rm int \, inel} \, . \label{causal}
\end{equation}
That the left-hand side is much smaller than the right-hand side reflects the extremely conservative nature of a constraint due to causality.   $\tau_i^{\rm int \, inel}$ would only be as small as $\frac{1}{\sqrt{4 B_i \mu_i}}/\gamma$ if the state were assembled at the speed of light.  However, weakly-bound states are highly nonrelativistic  and the natural speeds with which the constituents arrange themselves is much less than $c$.  Combining Inequalities (\ref{causal})  and  (\ref{ineq1}) implies that in order for weakly bound nuclei to form a nearly ideal gas in equilibrium with the hadrons in the gas (as assumed by the SHM) the following condition needs to be satisfied: 
\begin{equation}
\frac{m_i n_i}{\epsilon_i\sqrt{4 B_i \mu_i}}  \ll \tau_i^{\rm int \, inel}  \ll \frac{\tau_i}{2}\, . \label{ineq3}
\end{equation}

For a gas at $T=156.5$ MeV, the chemical freeze-out temperature of the SHM as fitted from LHC data, the constraints of Inequality (\ref{ineq3}) implies that for the SHM to be self-consistent for deuterons and hypertritons, $\tau_D^{\rm int \, inel} $ and $\tau_{{}_\Lambda^3 \rm H}^{\rm int \, inel} $ 
\begin{equation}
\begin{split}
2.7  \, {\rm fm}\, &  \ll \tau_D^{\rm int \, inel}  \ll  0.6 \, {\rm fm}\,  ,\\
5.4 \, {\rm fm}\, &  \ll \tau_{{}_\Lambda^3 \rm H}^{\rm int \, inel}  \ll 0.5 \, {\rm fm}\,,
\end{split}
\label{cons} \end{equation}
where the right-hand side of the inequality comes from the analysis of Sect.~\ref{taui} as given in Inequality (\ref{taunuc}) and the left-hand side is obtained from  $n_i$ and $\epsilon_i$ from Eqs.~(\ref{ni}) and (\ref{epsi}) along  known hadronic masses.

The obvious point is that Inequalities (\ref{cons}) cannot be satisfied for either the deuteron or the hypertriton.  In effect,  in a putative equilibrated hadron gas a would-be deuteron (hypertriton)  would be destroyed before it could be fully created; an equilibrated nearly ideal gas containing deuterons and  hypertritons simply cannot form.  The underlying reason is that the pions are sufficiently dense in such a medium that they will convert nucleons into $\Delta$s, destroying the would-be deuteron more rapidly than the deuteron could be assembled (or even have its physical configuration in causal contact).  Clearly this indicates that at least as far as the weakly-bound nuclei are concerned the assumptions underlying the statistical hadronization model are not consistent with the temperature extracted with the SHM from the LHC data.

It is worth noting that the inconsistency in the assumptions of the SHM for weakly-bound nuclei is almost certainly  much worse than indicated by  Inequalities (\ref{cons}), which are based on numerous conservative assumptions.  These assumptions render the left-hand sides of Inequalities (\ref{cons}) larger than the values quoted and simultaneously render the right-hand sides smaller.   Both of these act to worsen the inconsistencies.

 For example, we took the size of the state $R$ to be  $1/\sqrt{4 B \mu}$ which both neglected the size of the nucleon in the size of the nucleus and the effect of the short-range potential increasing the rms separation; using a realistic phenomenological potential would increase the left-hand side of the deuteron expression by something like 25\%.  We used the more conservative estimate in order to reduce model dependence.  More significantly, the causality bound is extremely conservative.  It assumes that the nuclear state forms at the speed of light.  However, weakly-bound nuclear states when viewed as a composite of nucleons are  nonrelativistic---the characteristic momentum in their wave functions is given by $\sqrt{\mu B}$.  Thus it is natural to assume that these constituents assemble with a speed of order $\sqrt{{B}/{\mu}}$.  If one accepts this assumption, the inconsistency from Inequality (\ref{cons}) worsens significantly: the left-hand of the inequality increases by something on the scale of an order of magnitude for the deuteron  and  two orders of magnitude for the hypertriton.

The right-hand side also contains numerous conservative estimates.  For example, the bound on the lifetime of the deuteron came from a bound on the rate at which the nucleons in it were converted into $\Delta$s via collisions with pions, which in turn was obtained via detailed balance by a bound on the rate of $\Delta$ annihilation from spontaneous decays.  Every step in that chain neglects things that act to shorten the lifetime.  We neglected $\Delta$ decays induced by collisions; we neglected processes in which nucleons were converted to baryons other than $\Delta$s; we neglected processes in which the pions break up the deuteron into two nucleons while leaving the nucleons intact; and we neglected processes in which the deuteron was broken up by collisions with hadrons other than pions.  There is evidence, for example, that the interaction of the deuteron or hypertriton with nucleons and antinucleons in the putative gas is too significant to neglect at $T=156.5$ MeV:  at any given time, a nucleon in the deuteron or hypertriton is typically interacting with at least one  nucleon or antinucleon in the gas with an interaction energy greater than its binding energy. Moreover it is straightforward to see that at any given moment, the weakly bound nuclei have numerous hadrons inside of them: their volume  (which be may estimated conservatively as $\left (4 \mu B \right )^{-3/2}$) is much larger than the inverse of the density of hadrons.  At an intuitive level this makes it very hard to argue that these nuclei are isolated objects in the medium as the model assumes.

We will not pursue in any detail the various  avenues to determine the extent to which the  inconsistency in the assumptions of the SHM  for the weakly-bound nuclei is worse than indicated in Inequalities (\ref{cons}).  Doing so requires analysis that is at least somewhat model dependent. In contrast, the analysis required to obtain  Inequalities (\ref{cons}) required only information used as inputs to the SHM such as masses and branching ratios, an output from SHM (the temperature at chemical freeze-out), well-established single hadron properties such as resonance lifetimes, plus the very reasonable assumption that the nuclei were sufficiently weakly bound that pion scattering process off of the constituents were well-approximated by their free-space values. 

In any case, it is unnecessary to document how much worse the inconsistency is 
compared to Inequalities (\ref{cons}): As they stand, these inequalities are sufficient to show that  the weakly-bound nuclei cannot form a nearly ideal gas in equilibrium with hadrons at or above the chemical freeze-out temperature $T_{\rm cf} = 156.5$ MeV extracted from the SHM.

One can use Inequality (\ref{ineq3}) to test the consistency at various 
temperatures. For example, at $T_{\rm cf}=120$ MeV, Inequalities (\ref{cons}) would be,
\begin{equation}
\begin{split}
2.8  \, {\rm fm}\, &  \ll \tau_D^{\rm int \, inel}  \ll  1.0 \, {\rm fm}\,  ,\\
5.5 \, {\rm fm}\, &  \ll \tau_{{}_\Lambda^3 \rm H}^{\rm int \, inel}  \ll 0.8 \, {\rm fm}\,,
\end{split}
\label{const}
\end{equation}
and would still be not satisfied.

\section{Discussion}

Of course, it has been recognized previously that the weak binding and large size of the deuteron and hypertriton make it quite remarkable that their yields can be described by a chemical freeze-out mechanism as assumed in the statistical hadronization model \cite{Andronic:2017pug}. The success of the model in predicting these yields is notable given an intuitive sense that the binding and size of these nuclei could be in  tension with the assumptions underlying the model given that the extracted chemical freeze-out temperature is close to two orders of magnitude greater than the binding energy of the deuteron and approximately three orders of magnitude greater than the binding energy of the hypertriton. This paper puts this intuitive sense into a concrete form:  Inequalities (\ref{cons}, \ref{const}) show that  for the weakly-bound nuclei, the assumption of chemical freeze-out from a near-ideal gas made in the SHM cannot be consistent  with the output of the model, a freeze-out temperature of $156.5$ MeV.

This result raises some major questions about the SHM.

One critical question is whether the yields of the hadrons (and more tightly bound nuclei) are also due to a  mechanism other than the chemical freeze-out assumed by the SHM.  The question is of importance since, as noted in the introduction, if one accepts the picture of a chemical freeze-out with the  temperature as given in the model, then one has evidence for a remarkable physical picture:  at sufficiently high energies the system essentially hadronizes, has all hadronic species equilibrate and then freezes out chemically before it cools noticeably below the hadronization temperature.  This  requires the temperature characterizing  an equilibrium thermodynamic quantity---the crossover from a quark-gluon regime to a hadronic regime---to be very nearly equal to the temperature set by a non-equilibrium physics fixed by the large-scale dynamics of the collision: the chemical freeze-out temperature.  
 
The fact that the yields of the weakly-bound nuclei are well-described by the model despite them not arising from chemical freeze-out shows that the mere fact that the model describes the yields cannot be taken as strong evidence that a chemical freeze-out occurs.  Moreover,  the fact that the model works for two distinct  species of weakly-bound nuclei that cannot be described via chemical freeze-out  mechanism makes the issue particularly acute.  Of course, it is logically possible that the yields of the hadrons (and perhaps more tightly-bound nuclei) are due to chemical freeze-out, while  the yield of one species of weakly-bound nuclei accidentally matches the model prediction due to an entirely unrelated mechanism;  however, it seems quite implausible for this to happen accidentally for two species.  This makes the possibility that the success of the SHM could be due to a mechanism other than the chemical freeze-out for all species worth taking seriously.  

We note that the SHM is not the only model proposed that describe the yields of hadrons;  coalescence models are based on radically different assumptions and also do a reasonable job in explaining the yields of hadrons \cite{Cho:2017dcy}.  The result in this paper suggests that whatever advantage the SHM has over coalescence models in describing yields of light nuclei should be treated quite cautiously as an argument in favor of the SHM more generally: at least with regard to the weakly-bound nuclei the success of SHM in describing yields cannot be ascribed to the mechanism upon which the model is åsupposed to be based.

Clearly, a key issue in light of  Inequalities (\ref{cons}, \ref{const}) is to  understand why the model  is able to describe the yields of the weakly-bound nuclei well despite the inconsistency with the model assumptions.  Reference \cite{Andronic:2017pug} speculates about the possibility that at QGP hadronization,  compact colorless objects with the quantum numbers of the weakly-bound nuclei could be produced that will eventually evolve into the nuclei.  It is argued that if these objects have a lifetime longer than about 5 fm and an excitation energy of around 40 MeV or less, they could account for the yields.  The 40 MeV excitation energy means that the mass of this object is close to the nuclear mass and would give similar yields to what is computed  in the SHM.  The compact size and 5 fm lifetime  are designed to ensure that it would survive in a hadronic phase before the phase undergoes chemical freeze-out.  

Unfortunately such an explanation appears to be untenable.    It is conceivable that in a dense medium the effective interactions between constituents of a bound state are modified by the medium and that this could lead to objects that are much more compact than their free-space analogs.  It is also conceivable that if the medium were to become more dilute sufficiently slowly, such an object might adiabatically evolve into its free-space analog--in this case a weakly-bound nucleus.  However, by construction, once the system enters the hadronic regime the interactions are the standard hadronic ones.  To the extent that such an object survives as a single compact state into the hadronic regime it will not evolve adiabatically into a weakly-bound nuclear state \cite{Shuryak:2018lgd}.  Rather, once in the hadronic regime the compact object can be expressed as a superposition of ordinary hadronic states.   As the state evolves in time the various hadronic components will decohere  and the system is left with various probabilities for obtaining the physical hadronic states.  To the extent that the state is very compact, its overlap with the nuclear bound state of interest (as opposed to multi-nucleon scattering states) is very small and thus so is the probability that the compact object will evolve into the nucleus of interest.  Thus, if  compact objects with the quantum number of weakly-bound nuclei were to form, the SHM would be expected to over-predict  the yields rather than getting them correct.  

For example, in the hadronic phase a compact object in its rest frame with the  quantum numbers of the deuteron and spin projection $m$   can be written as a superposition of a deuteron state  with amplitude $\alpha$ and a continuum two-nucleon state with amplitude $\beta$: 
\begin{equation}
\begin{split}
| {\rm compact \, object, m}\, \rangle & =\alpha |D , m \rangle + \beta |\psi_{\rm continuum} , m \rangle \\
 {\rm with} & \; |\alpha|^2+ |\beta|^2=1
\end{split}
\end{equation}
$|\psi_{\rm continuum} , m \rangle$ itself contains both s-wave an d-wave continuum contributions
\begin{equation}
\begin{split}
|\psi_{\rm continuum} , m \rangle &  =  \int \frac{dp}{2 \pi} \left (  f_0 (p) |p ,m \rangle_0 +  f_2 (p) |p , m  \rangle_2 \right )\\
{\rm with}  & \int \frac{dp}{2 \pi}\left ( |f_0|^2 +  |f_2 (p)|^2 \right ) =1
\end{split}
\end{equation}
where $|p ,m \rangle_0 $ are the appropriately normalized continuum scattering states for a proton and a neutron with net angular momentum 1 and spin projection $m$ whose angular dependence is either $L=0$ or $L=2$.  Note that in principle such  a compact object could contain other components such as pions, but given the constraint that the total energy is supposed to be  no more than 40 MeV or so above the deuteron such components are small.  Note that an analogous expression for the hypertriton is more involved: it contains many continuum channels since it has both a D-$\Lambda$ continuum and a three-body continuum with several possible channels.

The important point in the preceding example is that the probability that the state evolves into a deuteron is $P_D=|\alpha|^2 =| \langle D, m|{\rm compact \, object, m}\, \rangle|^2$.  The quantum state for a compact object has to be much different from that of the deuteron---otherwise it would not be compact and the mechanism would fail.  Thus one expects for the mechanism  to work $| \langle D, m|{\rm compact \, object, m}\, \rangle|^2 \ll 1$ and the deuteron yields to be much less than predicted by the model and similarly for the hypertriton yields.  Indeed for the hypertriton the overlap should be exceptionally small since the hypertriton state is so large.

To conclude, the analysis in this paper shows conclusively that the yields of weakly-bound light nuclei are not due to a chemical freeze-out from an equilibrated nearly ideal hadronic gas as is assumed in the SHM, despite the fact that the SHM describes these yields well.  At this point it is an open question as to why the SHM is able to describe these yields.  It is also an open question as whether the yields of hadrons should be understood to be due to chemical freeze-out or by some other mechanism.

\begin{acknowledgments}
The authors would like to thank Ralf Rapp, Fuqiang Wang, Xiaofeng Luo, Stanislaw Mrowczynski, Jinhui Chen, and Dmytro Oliinychenko for helpful discussions. Y.C., T.D.C., and Y.Y. are supported by the U.S. Department of Energy under Contract No. DE-FG02-93ER-40762. B.A.G. is grateful for the hospitality of the Maryland Center for Fundamental Physics at the University of Maryland, College Park, and the Department of Applied Mathematics and Theoretical Physics at the University of Cambridge, UK, where this work was partially done. B.A.G. was supported in part by the PSC‐CUNY Award No. 69428‐00 47. 

\end{acknowledgments}

\bibliographystyle{apsrev4-1}
\bibliography{shm}

\end{document}